\documentclass[prd,twocolumn,aps,superscriptaddress]{revtex4}
\usepackage{amsfonts}
\usepackage{mathrsfs}
\usepackage{amsmath}% needed for subequations
\usepackage{color}
\usepackage{graphicx}
\usepackage{bm}% bold maths
\usepackage{amssymb}
\usepackage{xspace}
\usepackage{epstopdf}
\usepackage{dcolumn}% Align table columns on decimal point
\usepackage{longtable}
\usepackage{multirow}
\usepackage[colorlinks=true, letterpaper=true, pdfstartview=FitV, linkcolor=blue, citecolor=blue, urlcolor=blue]{hyperref}

\begin{document}
\title{Weyl-loop half metal in Li$_3$(FeO$_3$)$_2$}

\author {Cong Chen}
\affiliation{Department of Physics, Key Laboratory of Micro-nano Measurement-Manipulation and Physics (Ministry of Education), Beihang University, Beijing 100191, China}
\affiliation{Research Laboratory for Quantum Materials, Singapore University of Technology and Design, Singapore 487372, Singapore}

\author{Zhi-Ming Yu}
\affiliation{Research Laboratory for Quantum Materials, Singapore University of Technology and Design, Singapore 487372, Singapore}

\author{Si Li}
\affiliation{Research Laboratory for Quantum Materials, Singapore University of Technology and Design, Singapore 487372, Singapore}

\author {Ziyu Chen}
\affiliation{Department of Physics, Key Laboratory of Micro-nano Measurement-Manipulation and Physics (Ministry of Education), Beihang University, Beijing 100191, China}

\author {Xian-Lei Sheng}
\email{xlsheng@buaa.edu.cn}
\affiliation{Department of Physics, Key Laboratory of Micro-nano Measurement-Manipulation and Physics (Ministry of Education), Beihang University, Beijing 100191, China}
\affiliation{Research Laboratory for Quantum Materials, Singapore University of Technology and Design, Singapore 487372, Singapore}

\author{Shengyuan A. Yang}
\affiliation{Research Laboratory for Quantum Materials, Singapore University of Technology and Design, Singapore 487372, Singapore}
\affiliation{Center for Quantum Transport and Thermal Energy Science, School of Physics and Technology, Nanjing Normal University, Nanjing 210023, China}

\begin{abstract}
Nodal-line metals and semimetals, as interesting topological states of matter, have been mostly studied in nonmagnetic materials. Here, based on first-principles calculations and symmetry analysis, we predict that fully spin-polarized Weyl loops can be realized in the half metal state for the three-dimensional material Li$_3$(FeO$_3$)$_2$. We show that this material has a ferromagnetic ground state, and it is a half metal with only a single spin channel present near the Fermi level. The spin-up bands form two separate Weyl loops close to the Fermi level, which arise from band inversions and are protected by the glide mirror symmetry. One loop is of type-I, whereas the other loop is of hybrid type. Corresponding to these two loops in the bulk, on the (100) surface, there exist two fully spin-polarized drumheads of surface states within the surface projections of the loops. The effects of the electron correlation and the spin-orbit coupling, as well as the possible hourglass Weyl chains in the nonmagnetic state have been discussed. The realization of fully spin-polarized Weyl-loop fermions in the bulk and drumhead fermions on the surface for a half metal may generate promising applications in spintronics.
\end{abstract}

\maketitle

\section{Introduction}

Topological metals and semimetals have been attracting extensive attention in recent research~\cite{Chiu2016,Burkov2016,Yang2016,Dai2016,Bansil2016}. In these materials, the electronic band structure exhibits protected band crossings near the Fermi level. The low-energy electrons around these band crossings may acquire an emergent pseudospin degree of freedom and can have distinct types of dispersions.
For example, in the Weyl/Dirac semimetals, the conduction and valence bands cross at isolated twofold/fourfold degenerate nodal points
in the Brillouin zone (BZ), such that the low-energy electrons resemble the relativistic Weyl/Dirac fermions~\cite{Wan2011,Murakami2007,Burkov2011,Young2012,Wang2012b,Wang2013b,Zhao2013c,Yang2014a,Liu2014c,Borisenko2014,Weng2015,Lv2015,Xu2015a,chen2017,Vishwanath2018},
and hence many fascinating phenomena in relativity and high-energy physics may be simulated in condensed matter experiments~\cite{Nielsen1983,Son2013,Guan2017up}.

For a three-dimensional (3D) crystalline material, besides the 0D nodal points, the band crossing may also form 1D nodal lines~\cite{Fang2016} or 2D nodal surfaces~\cite{Zhong2016,Liang2016,WuWK2018}. A variety of nodal lines have been discovered~\cite{Yang2014,Weng2015c,Mullen2015,Yu2015,Kim2015a,Chen2015,Fang2015,Chen2015a,Li2016,Bian2016,Schoop2016,Gan,Huang2016}, and they may be classified according to different characteristics. For example, a single nodal loop can be classified by its winding pattern around the BZ (which is a three-torus)~\cite{Li2017}, characterized by a $\mathbb{Z}^3$ index, so a loop circling around a high-symmetry point is distinct from a loop traversing the BZ. A nodal line can also be classified according to the type of energy dispersion, leading to type-I, type-II~\cite{Li2017}, and hybrid nodal lines~\cite{Li2017,ZhangXM2018}. The classification can also be based on the order of dispersion: besides the conventional linear order dispersion, Yu \emph{et al.}~\cite{Yu2018q} have shown that quadratic and cubic nodal lines can also exist. In addition, when two or more lines are present, they may interconnect and form interesting patterns in the BZ, such as crossing nodal rings~\cite{Weng2015c,Yu2015,YuWC2018}, nodal boxes~\cite{Sheng2017JPCL}, nodal chains~\cite{Bzdusek2016,Wang2017,Yu2017,Yan2017a}, and Hopf links~\cite{Zhong2017NC,ChenW2017,YanZB2017,ChangGQ2017,ZhouY2018}.

The previous study on nodal lines are mostly in nonmagnetic materials, whereas the examples with magnetic nodal lines are rather limited, such as MnF$_3$~\cite{Jiao2017}, some cubic magnetic oxides~\cite{Wang2018b}, and magnetically ordered GdSbTe~\cite{Hosen2018ti}. Two classes of nodal lines in antiferromagnetic systems have been proposed by model analysis~\cite{WangJ2017}. Obviously, the nodal lines in magnetic materials represent another distinct type. Particularly, it would be most interesting when the material is a half metal, i.e., when the bands around the Fermi level are all belong to a single spin channel, then the nodal line would be completely spin polarized. Consequently, all the interesting physical properties associated with the nodal-line fermions happen only for a single spin, and the spin polarization can be switched by controlling the magnetization direction. Thus, such topological half metals have great potential for spintronics applications for information storage and processing, with the advantages of high speed and low power consumption.

In this work, by first-principles calculations,  we reveal the 3D material Li$_3$(FeO$_3$)$_2$ as a Weyl-loop half metal with fully spin-polarized topological features.
We show that the material has a ferromagnetic ground state, and it is a half metal: only spin-up bands are present around the Fermi level, whereas the spin-down bands have an energy gap $\sim 1.35$ eV. Remarkably, the crossing between the spin-up bands form two separate doubly degenerate Weyl loops close to the Fermi level. One Weyl loop centered around the $R$ point in the $k_x=\pi$ plane is a type-I loop, while the other centered around the $\Gamma$ point in the $k_x=0$ plane is a hybrid loop. Such pattern of magnetic Weyl loops has not been reported before. Corresponding to these two loops, there exist fully spin-polarized topological drumhead surface states in two separate regions on the (100) surface. The robustness of the band features against spin-orbit coupling (SOC) and correlation effects are discussed. In addition, we show that by substituting Fe with Te, the material can become a metal with Weyl chains dictated by nonsymmorphic symmetries. Our work reveals a new magnetic topological state with fully spin-polarized Weyl loops and drumhead surface states, which may have promising applications in spintronics.

\section{Computational Method}
\label{methods}

The first-principles calculations are based on the density-functional theory (DFT) as implemented in the Vienna \textit{ab initio} simulation package (VASP)~\cite{Kresse1994,Kresse1996}, using the projector augmented wave method~\cite{PAW}. The generalized gradient approximation (GGA) with Perdew-Burke-Ernzerhof (PBE)~\cite{PBE} realization was adopted for the exchange-correlation potential. The plane-wave cutoff energy was set to 500 eV. The Monkhorst-Pack $k$-point mesh~\cite{PhysRevB.13.5188} of size $11\times11\times 7$ was used for the BZ sampling.  To account for the correlation effects for transition metal elements, the DFT$+U$ method~\cite{Anisimov1991,dudarev1998} was used for calculating the band structures. $U$ values between $0$ and $6$ eV have been tested for the Fe-$3d$ orbitals. The crystal structure was optimized until the forces on the ions were less than 0.01 eV/\AA. From the DFT results, the maximally localized Wannier functions (MLWF) for  Fe-$3d$ and O-$2p$ orbitals were constructed, based on which the tight-binding models for bulk and semi-infinite systems were developed to study the surface states~\cite{Marzari1997,Souza2001,Wu2017,Green}.

\section{Crystal Structure}

\begin{figure}[t!]
\includegraphics[width=8cm]{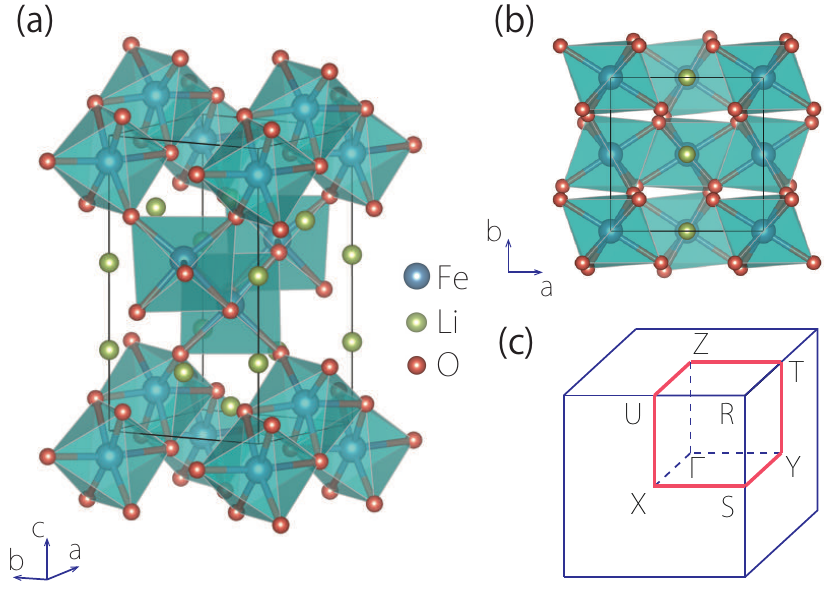}
\caption{(a) Side view and (b) top view of the crystal structure for Li$_3$(FeO$_3$)$_2$. (c) The corresponding Brillouin zone. The red lines indicate the paths where the bands are twofold degenerate (see the discussion in Sec.~V).}
\label{stru}
\end{figure}

\begin{table}[b!]
   \caption{Wyckoff positions for the nonequivalent atoms in  Li$_3$(FeO$_3$)$_2$.}
   \begin{tabular}{cccccccc}
      \hline  \hline
      & Atom  & Wyckoff letter & $x$      & $y $      & $z$ \\
      \hline
      & Li$_1$     & 2a  & 0.5     & 0.5     & 0.7739  \\
      & Li$_2$     & 2a  & 0.5     & 0.5     & 0.0694  \\
      & Li$_3$   & 2b  & 0.5     & 0.0     & 0.7208  \\
      & Fe$_1$    & 2a  & 0.5     & 0.5     & 0.4132  \\
      & Fe$_2$    & 2b  & 0.5     & 0.0     & 0.1015  \\
      & O$_1$    & 4c  & 0.3507  & 0.7421  & 0.2475  \\
      & O$_2$    & 4c  & 0.3289  & 0.7949  & 0.9272  \\
      & O$_3$    & 4c  & 0.3146  & 0.7070  & 0.5867  \\
      \hline   \hline
    \end{tabular}
    \label{tab1}
\end{table}

The crystal lattice structure for Li$_3$(FeO$_3$)$_2$ is shown in Fig.~\ref{stru}(a-b), which belongs to the orthorhombic crystal system, with space group No.~34 ($Pnn2$). The lattice consists of a framework of Fe atoms, with each Fe atom surrounded by six oxygen atoms, forming an octahedral crystal field, and the Li atoms are intercalated into the framework to reach electronic neutrality. One unit cell here contains two formula units of Li$_3$(FeO$_3$)$_2$. The Wyckoff coordinates are given in Table~\ref{tab1}. This structure was proposed in Materials Project~\cite{Jain2013}, which has been demonstrated to be energetically and dynamically stable. The optimized lattice constants are $a_0=4.7952$ \AA , $b_0=4.8122$ \AA, and $c_0=8.1250$ \AA. There are two glide mirror planes involving half lattice translations $\mathcal{\tilde{M}}_x: (x, y, z) \rightarrow (-x + \frac{1}{2}, y+\frac{1}{2},z+\frac{1}{2})$ and $\mathcal{\tilde{M}}_y: (x,y,z) \rightarrow (x+\frac{1}{2},-y+\frac{1}{2},z+\frac{1}{2})$. Combining these two operations leads to the twofold  rotation $C_{2z}$. These symmetries will be important for our discussion below.

\section{Magnetic configuration}

\begin{figure}[b!]
\includegraphics[width=8.5 cm]{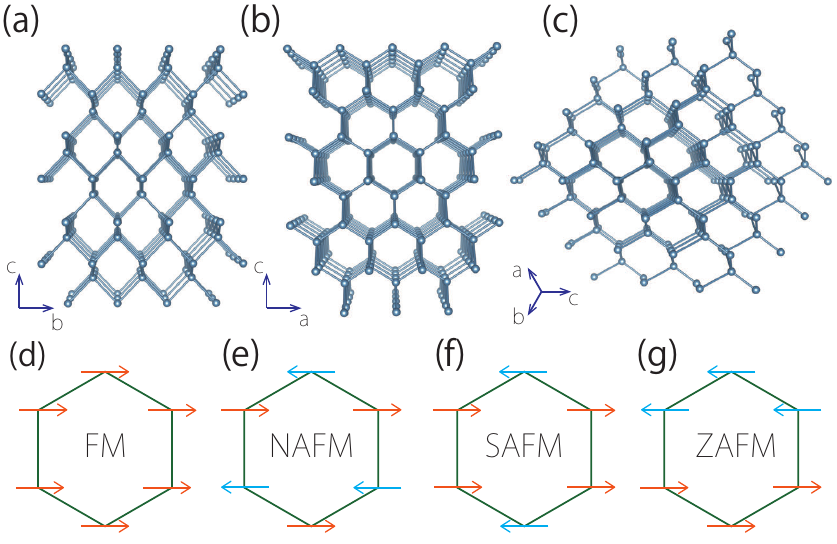}
\caption{Perspective views of a $3\times 3$ super-cell for the Fe framework, from (a) [100], (b) [010], and (c) [111] directions. The lower panels illustrate the possible magnetic configurations that we have considered: (d) ferromagnetism (FM), (e) N\'{e}el antiferromagnetism (NAFM), (f) striped AFM (SAFM), and (g) zigzag AFM (ZAFM).}
\label{mag}
\end{figure}

Before studying electronic structure, it is important to determine the magnetic ground state for the material. The $3d$ element Fe in Li$_3$(FeO$_3$)$_2$ carries a nonzero magnetic moment. These moments spontaneously order in the ground state.
To facilitate the study of magnetic configurations, here we focus on the framework consisting of only the Fe atoms. Figures~\ref{mag}(a-c) illustrate
the framework of Fe atoms, which displays honeycomb-like structure when viewed from different directions. Here, we investigate four different kinds of possible magnetic configurations as shown in Figs.~\ref{mag}(d-g). These include the ferromagnetic (FM) state [Fig.~\ref{mag}(d)], the N\'{e}el-type antiferromagnetic (NAFM) state, the striped antiferromagnetic (SAFM) state, and the zigzag-type antiferromagnetic (ZAFM) state. For each of these states, we consider four possible orientations for the magnetic moments, namely, the [100], the [010], the [001], and the [111] directions.

\begin{table}[b]
  \caption{Total energy $E_\mathrm{tot}$ per unit cell (in eV, relative to that of the FM$_{[111]}$ ground state),
    as well as magnetic moment $M$ (in $\mu B$) per Fe atom, obtained for several magnetic configurations as illustrated in Fig.~\ref{mag}. The values are calculated by GGA+SOC method with $U=4.0$ eV. Paramagnetic state has
    $E_\mathrm{tot}=6.660$ eV and $M=0$.}\label{tab:magnet}
  \begin{tabular}{cccccccc}
    \hline  \hline
                         & FM$_{[111]}$ & FM$^x$ & FM$^y$ & FM$^z$  &NAFM      &SAFM      &ZAFM \\
    \hline
    $E_\mathrm{tot}$     &0.0         &3.450   &3.463   &3.456    &1.017     &2.719     &2.714  \\
    $M$                    &3.504       &1.550   &1.617   &1.616    &2.983     &2.048     &2.133  \\
    \hline  \hline
  \end{tabular}
\end{table}

Our first-principles calculations show that  the lowest energy occurs for the
FM$_{[111]}$ configuration, i.e., when the magnetic moments are aligned in the [111] direction (pointing from one Fe atom to another equivalent Fe atom). The comparison of the total energies for these states are shown in Table~\ref{tab:magnet}. In the table, the values shown for the AFM states are for the respective lowest energy spin orientations. The large energy difference between the FM and the nonmagnetic states indicates the high stability of the magnetic ordering. In the following, we take FM$_{[111]}$ as the ground state configuration for studying the electronic band structure for Li$_3$(FeO$_3$)$_2$.

\section{spin-polarized Weyl loops}

\begin{figure}[t!]
\includegraphics[width=8cm]{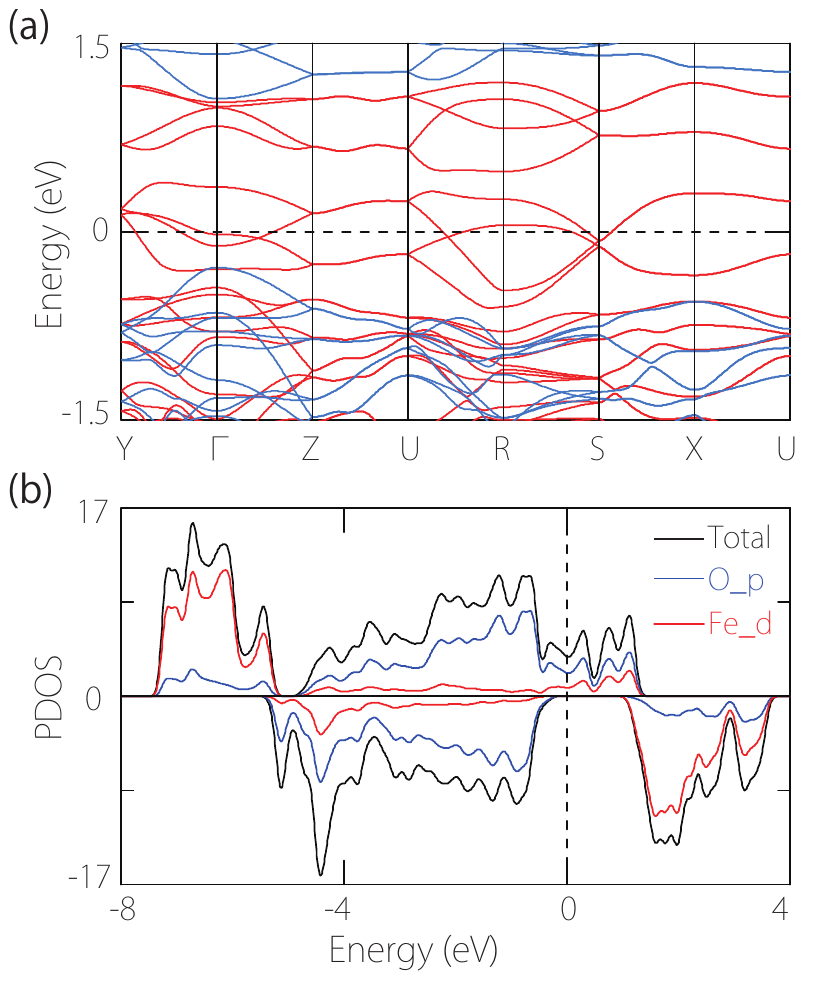}
\caption{(a) Band structure of Li$_3$(FeO$_3$)$_2$ with on-site Coulomb repulsion $U = 4.0$ eV (red and blue lines correspond to spin-up and spin-down channels, respectively). (b) Spin-polarized total and projected density of states of Li$_3$(FeO$_3$)$_2$, with the spin-up taking positive values and the spin-down taking negative values.}
\label{band}
\end{figure}

\begin{figure}[t!]
\includegraphics[width=8cm]{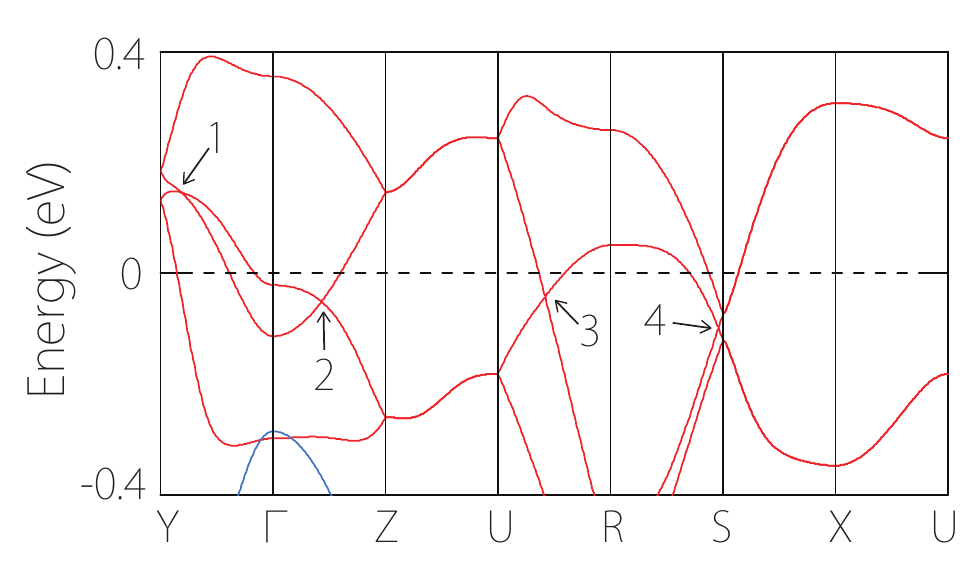}
\caption{Enlarged view of the band structure of Li$_3$(FeO$_3$)$_2$ near the Fermi level. The band-crossing points are labeled by the numbers.}
\label{bandzoom}
\end{figure}

After identifying the magnetic ground state, we then concentrate on the electronic band structure. Figure~\ref{band}(a) shows its detailed band structure for  Li$_3$(FeO$_3$)$_2$ in the FM$_{[111]}$ configuration. Here, the $U$ value is taken to be 4 eV, which is typical for Fe. SOC is neglected for now, because its strength is small for the light elements involved in this material. The effects of the $U$ value and the SOC on the band features will be discussed later.

One observes that the material is a half metal, with one channel (spin up) being metallic and another channel (spin down) being insulating. From the projected density of states (PDOS) as displayed in Fig.~\ref{band}(b), one clearly sees that
the states around the Fermi energy are fully polarized in the spin-up channel, while the spin-down channel has a large gap of about 1.35 eV. The low-energy states are dominated by the Fe-3$d$ and the O-2$p$ orbitals.

In Fig.~\ref{bandzoom}, we show an enlarged view of the low-energy bands around the Fermi level. Two kinds of features can be observed. First, the bands are doubly degenerate along the high-symmetry paths indicated by the red lines in Fig.~\ref{stru}(c) (some are not shown in Fig.~\ref{bandzoom}). Second, linear band crossing points appear on the paths $\Gamma$-$Y$, $\Gamma$-$Z$, $R$-$U$, and $R$-$S$, as indicated in Fig.~\ref{bandzoom}.

Let's first investigate the double degeneracy appearing on the high-symmetry paths marked in Fig.~\ref{stru}(c). We shall show that such degeneracy is guaranteed by the symmetry of the system.

Before proceeding, it is important to note that without SOC, the spin and the orbital degrees of freedom are independent and can be regarded as different subspaces. The spin orientation does not affect the orbital part of the wave function. With a chosen spin polarization axis,
the two spin channels are decoupled, and hence in terms of symmetry properties, the bands for each spin species can be \emph{effectively} regarded as for a \emph{spinless} system. Thus, regarding the states of one spin, such as the spin-up bands here, all the crystalline symmetries are preserved~\cite{Wang2016L,ChangGQ2017}.

For example, consider the degeneracy along the $U$-$X$ path. One notes that any $k$ point on this path is invariant under both $\mathcal{\tilde{M}}_x$ and $\mathcal{\tilde{M}}_y$. The commutation relationship between $\mathcal{\tilde{M}}_x$ and $\mathcal{\tilde{M}}_y$ is given by
\begin{equation}\label{Eq1}
\mathcal{\tilde{M}}_x \mathcal{\tilde{M}}_y = T_{\bar{1}10} \mathcal{\tilde{M}}_y \mathcal{\tilde{M}}_x,
\end{equation}
where $T_{\bar{1}10}=e^{ik_x-ik_y}$ represents the translation along the $[\bar{1}10]$ direction by one unit cell. Along $U$-$X$, we have $k_x=\pi$ and $k_y=0$, hence $T_{\bar{1}10}=-1$. Therefore, the two glide mirrors anti-commutate along this path. As a result, for any energy eigenstate $|u\rangle$ with $\mathcal{\tilde{M}}_x$ eigenvalue $g_x$, it must have a degenerate partner $\mathcal{\tilde{M}}_y|u\rangle$ with $\mathcal{\tilde{M}}_x$ eigenvalue $-g_x$. This proves that the double degeneracy on $U$-$X$ is guaranteed by symmetry.

\begin{figure}[t!]
\includegraphics[width=8 cm]{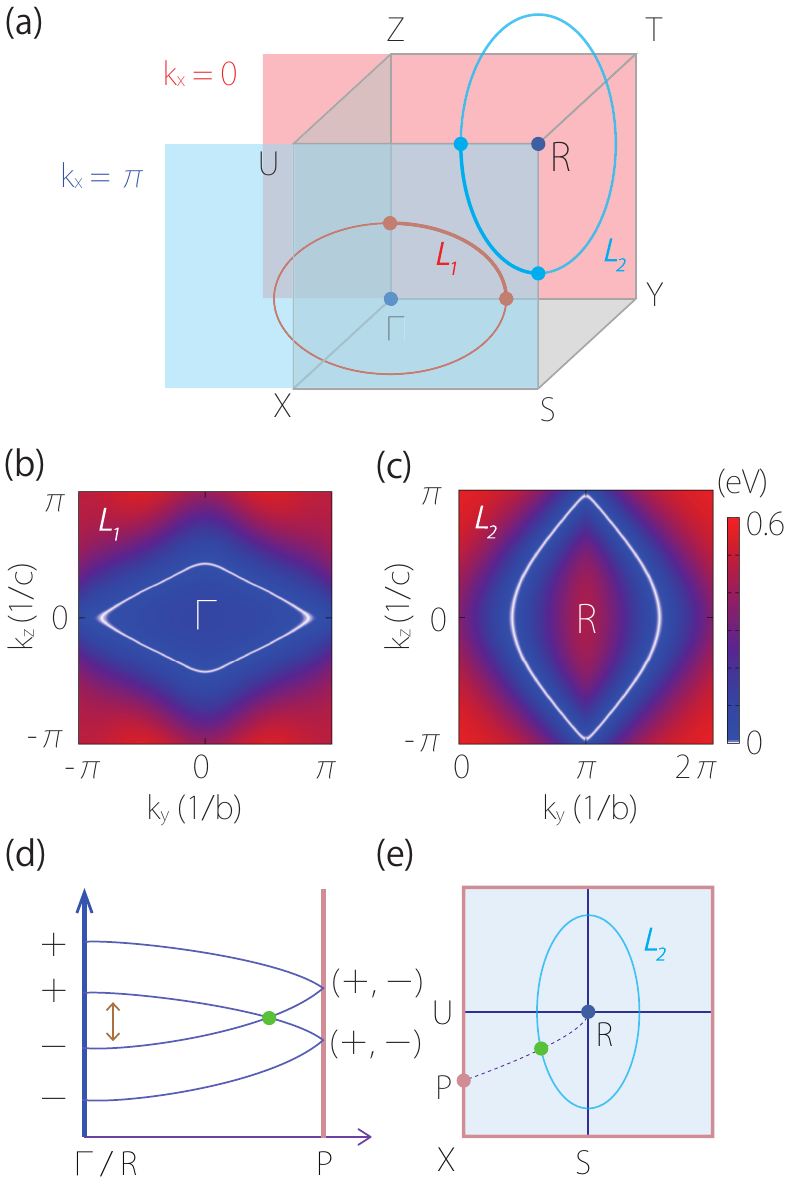}
\caption{(a) Schematic figure showing the two fully spin-polarized Weyl loops: loop $L_1$ in the $k_x = 0$ plane centered at $\Gamma$ and loop $L_2$ in the $k_x = \pi$ plane centered at $R$. This figure shows a reduced BZ. (b-c) Shape of the two Weyl loops obtained from DFT calculations. The color map indicates the local gap between the two crossing bands. (d) Schematic figure of the band ordering along a generic path connecting $\Gamma$ (or $R$) to a point $P $ on Z-T-Y (or U-X-S). The labels indicate the  $\mathcal{\tilde{M}}_x$ eigenvalues. (e) The crossing point in (d) (denoted by the green dot) traces out the Weyl loop on the $k_x=\pi$ (or $k_x=0$) plane.}
\label{nodal}
\end{figure}

Next, we consider the path $U$-$Z$ with $k_y = 0$ and $k_z = \pi$. It is an invariant subspace for the combined operation $ \mathcal{T}\mathcal{\tilde{M}}_x $. We note that
\begin{equation}
(\mathcal{T\tilde{M}}_x )^{2} = T_{011}=e^{-i k_y - i k_z},
\end{equation}
where we have used $\mathcal{T}^2=1$ for the spinless case. Consequently, $(\mathcal{T}\mathcal{\tilde{M}}_x )^2=-1$ for any $k$ point on $U$-$Z$. This anti-unitary operator thus generates a Kramers-like double degeneracy on on $U$-$Z$.

Similar analysis applies for the other four paths $Z$-$T$, $T$-$Y$, $Y$-$S$, and $S$-$X$, showing that all the bands are doubly degenerate along these paths.

Now, let's turn to investigate the band crossing points, as labeled by 1 to 4  in Fig.~\ref{bandzoom}. A careful scan of the band structure near these crossing points shows that these points are in fact not isolated. Instead, they are located on two separate nodal loops, as illustrated in Fig.~\ref{nodal}(a). One loop (denoted as $L_1$) lies in the $k_x=0$ plane, centered at the $\Gamma$ point. The other loop (denoted as $L_2$) lies in the $k_x=\pi$ plane, centered at the $R$ point. Figure~\ref{nodal}(b-c) shows the shape of the two nodal loops obtained from the first-principles calculations.

In the following, we show that the two loops are protected by the $\mathcal{\tilde{M}}_x $ symmetry and are caused by band inversion. Let's consider $L_2$ in the $k_x=\pi$ plane. Here, we take a path $\ell$ which connects $R$ to some arbitrary point $P$ on $U$-$X$. According to the previous discussion, each state at $P$ has a double degeneracy, and the degenerate partners have opposite $\mathcal{\tilde{M}}_x $ eigenvalues $\pm g_x$, which are labeled as $(+,-)$ in Fig.~\ref{nodal}(d). As schematically shown in Fig.~\ref{nodal}(d), at $P$, there are two such degenerate pairs (four states) near the Fermi level. On the other hand, the corresponding four states are not required to be degenerate at $R$, where the $\mathcal{\tilde{M}}_x $ eigenvalues are $(-,-,+,+)$ for the states in ascending order. Along the path $\ell$, the four bands form a pattern shown in Fig.~\ref{nodal}(d). Focusing on the middle two bands, they are of opposite $\mathcal{\tilde{M}}_x $  eigenvalues, and their ordering is inverted between $P$ and $R$. As a result, they must cross along $\ell$ and the crossing point is protected by $\mathcal{\tilde{M}}_x$. The analysis applies for an arbitrary path connecting $R$ to an arbitrary point on $U$-$X$ and also $X$-$S$. Thus, the crossing point will trace out a nodal loop on the $k_x=\pi$ point centered at $R$, protected by the $\mathcal{\tilde{M}}_x$ symmetry, as shown in Fig.~\ref{nodal}(e). Since the crossing is doubly degenerate and is linear, the loop $L_2$ is a Weyl loop. Similarly, another Weyl loop $L_1$ appears in $k_x=0$ plane, and is also symmetry-protected.

As we have mentioned, a nodal loop can be classified as type-I, type-II, or hybrid type~\cite{Li2017,ZhangXM2018}, based on the type of dispersion for the points on the loop. After careful scan of the dispersion around each loop, we find that the loop $L_1$ is a hybrid loop, composed of both type-I and type-II nodal points. On the other hand, the loop $L_2$ is type-I. In addition, since each loop is locally located around a high-symmetry point, not traversing the BZ, the corresponding $\mathbb{Z}^3$ index characterizing its winding in BZ is trivial~\cite{Li2017}.

The most important feature of the Weyl loops here is that they are fully spin polarized. The nodal loops occur in the ferromagnetic state. More importantly, they belong to a single spin channel, due to the crossing between spin-up bands. Therefore, the low-energy nodal-loop fermions are fully spin polarized, which will be useful for spintronics applications.

\section{Spin-polarized Drumhead surface states}

\begin{figure}[b!]
\includegraphics[width=8.5 cm]{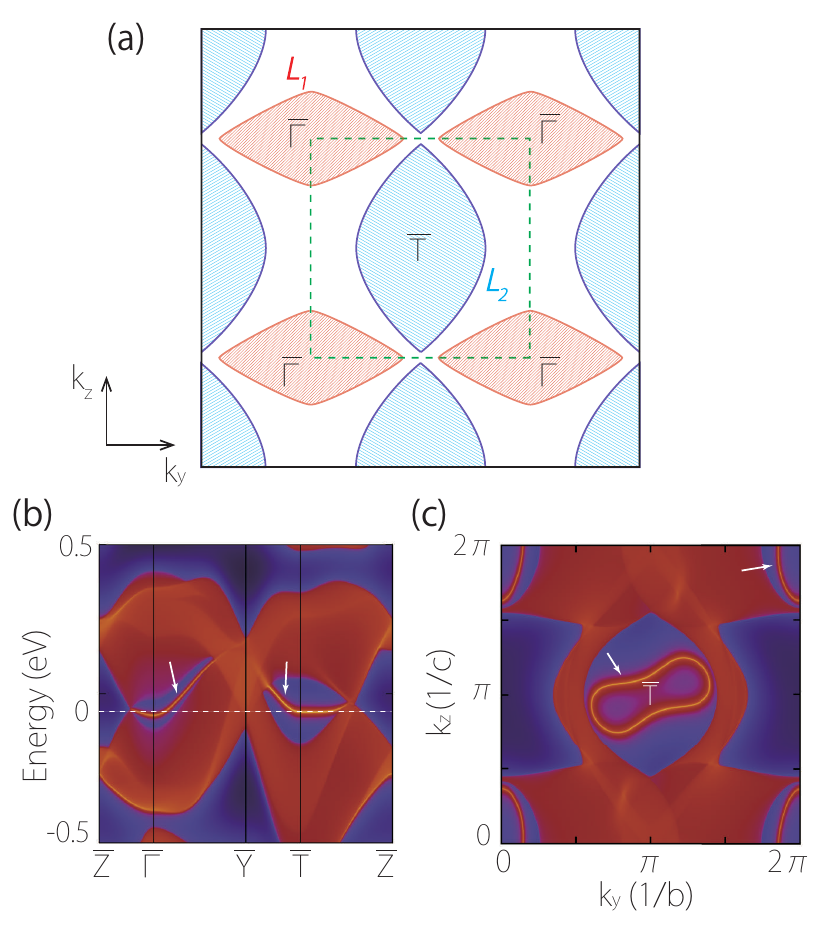}
\caption{Topological surface states. (a) Schematic view of the projections of the two bulk Weyl loops on the (100) surface. The red loop is from $L_1$ and the blue one is from $L_2$. The drumhead surface states are located inside these projected loops, indicated by the shaded regions. The dashed box here shows a unit surface BZ. (b) Projected spectrum on the (100) surface for Li$_3$(FeO$_3$)$_2$. (c) shows a constant energy slice at $-50.5$ meV. The arrows in (c) and (d) indicate the drumhead surface states. These states are also fully spin polarized.}
\label{surf}
\end{figure}

It is known that nodal loops in the bulk typically generates drumhead type surface states within the projected area of the loop in the surface BZ~\cite{Yang2014,Weng2015c}.
For Li$_3$(FeO$_3$)$_2$ studied here, since the bulk has fully spin polarized Weyl loops, one may expect that there would exist fully spin polarized drumhead surface states. Another interesting point is that there are two separate Weyl loops in the bulk. Then will they each produce a drumhead of surface states?

To address these issues, we have calculated the surface spectrum for the (100) surface, on which the two Weyl loops can have finite projected areas. Figure~\ref{surf}(b) shows the surface spectrum along high-symmetry paths in the surface BZ. One observes the drumhead type surface states around $\overline{\Gamma}$ and $\overline{T}$, as indicated by the arrows. These surface states indeed appear inside the projected loops. In Figure~\ref{surf}(a), we plot the projected loops in the surface BZ. The two loops do not overlap, and each has a finite projection. The surface states in Fig.~\ref{surf}(b) form two drumheads inside these two projected loops, marked by the shaded region in Fig.~\ref{surf}(a).
Figure~\ref{surf}(c) shows the constant energy slice at $E=-50.5$ meV, which cuts through the two drumheads (because they are not completely flat in energy), forming the two Fermi circles as indicated by the arrows. We have analyzed the spin polarization of these surface states, and confirm that they are fully polarized in the spin-up state.

\begin{figure}[b!]
\includegraphics[width=8 cm]{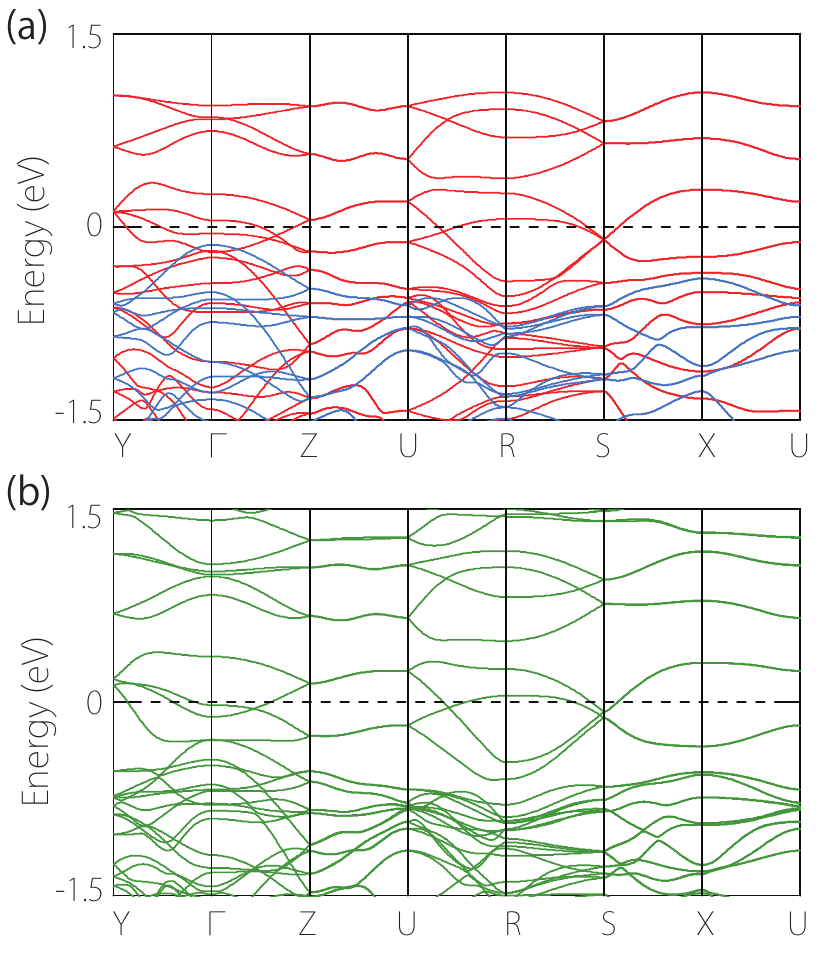}
\caption{ Band structure of Li$_3$(FeO$_3$)$_2$, (a) with $U=6.0$ eV (without SOC), and (b) with SOC and $U=4.0$ eV.}
\label{bandSOCU}
\end{figure}

\section{Discussion and conclusion}

We have demonstrated that Li$_3$(FeO$_3$)$_2$ is a half metal with two fully spin polarized Weyl loops near the Fermi level. In the calculation, we have set the Hubbard $U=4$ eV, a typical value for Fe 3$d$ orbitals. How robust is our result against the variation in $U$? To check this, we have tested $U$ values up to 6 eV. We find that the key band features such as the half metal character and the Weyl loops are robust against the variation in $U$. For instance, Figure~\ref{bandSOCU}(a) shows the band structure result with $U=6$ eV. One observes that the system is still a half metal, and the low-energy bands are very similar to the result in Fig.~\ref{band}(a). The main difference is that the gap for the spin-down channel is increased from 1.35 eV to 2.12 eV.

We have mentioned that the SOC effect is negligible for Li$_3$(FeO$_3$)$_2$, which consists of only the light elements. To verify this, we show the DFT band structure with SOC in Fig.~\ref{bandSOCU}(b). Compared to Fig.~\ref{band}(a), one can see that the bands are almost unaffected by SOC. Only when we zoom in the small region at the band crossing points, a small SOC gap can be observed. Our calculation shows that the SOC gap is very small ($<4$ meV) at the Weyl loops. Thus, the SOC effect is indeed negligible.

It has been demonstrated that even for cases with strong SOC, certain nonsymmorphic space group symmetries can protect interesting topological band features. For Li$_3$(FeO$_3$)$_2$, we already have two glide mirror planes. The SOC can be enhanced by replacing Fe by some heavier element such as Te. We have studied the resulting material Li$_3$(TeO$_3$)$_2$. This materials is non-magnetic and its band structure (with SOC) is shown in Fig.~\ref{TeSOCband}(a). From the zoom-in plot in Fig.~\ref{TeSOCband}(b-d), one can observe the hourglass type dispersions~\cite{Wang2016a,Young2015a}. The neck point in the hourglass dispersion traces out two Weyl chains in the BZ, as schematically shown in Fig.~\ref{TeSOCband}(e), dictated by the nonsymmorphic symmetries~\cite{Bzdusek2016,Wang2017}. However, the band splitting in Li$_3$(TeO$_3$)$_2$ is not large enough, so the Weyl chain features may not be easily resolved in experiment. Other candidate materials with the similar structure may be explored in the future.

In conclusion, we reveal that Li$_3$(FeO$_3$)$_2$ is a Weyl-loop half metal. We show that the material has a ferromagnetic ground state. It is metallic in one spin channel, yet insulating in the other spin channel. The low-energy bands form doubly degenerate lines along several high symmetry paths, and form two separate Weyl loops close to the Fermi level. The Weyl loops are in a single spin channel, hence they are fully spin polarized. We show that they lead to two drumheads of surface states on the (100) surface, which are also fully spin polarized. Such ferromagnetic Weyl loops and drumhead surface states may have great potential in spintronics applications.

\begin{acknowledgements}
The authors thank D. L. Deng for helpful discussions. This work is supported by the NSF of China (Grant No.~11504013, No.~61227902, No.~11474015 and No.~11774018), and the Singapore Ministry of Education AcRF Tier 2 (MOE2015-T2-2-144 and MOE2017-T2-2-108).
\end{acknowledgements}

\begin{appendix}

\renewcommand{\theequation}{A\arabic{equation}}
\setcounter{equation}{0}
\renewcommand{\thefigure}{A\arabic{figure}}
\setcounter{figure}{0}
\renewcommand{\thetable}{A\arabic{table}}
\setcounter{table}{0}

\section{Hourglass Weyl chain in Li$_3$(TeO$_3$)$_2$}

Here, we present a symmetry analysis for the Weyl chains in Li$_3$(TeO$_3$)$_2$.

%\begin{figure}[t!]
%\includegraphics[width=8.5 cm]{Tebandhalf.pdf}
%\caption{(a) Band structure of Li$_3$(TeO$_3$)$_2$ in the absence of  SOC. (b) Projected density of states (PDOS). (c) Schematic figure showing the nodal-box. }
%\label{Tebandhalf}
%\end{figure}

\begin{figure}[t!]
\includegraphics[width=8.5 cm]{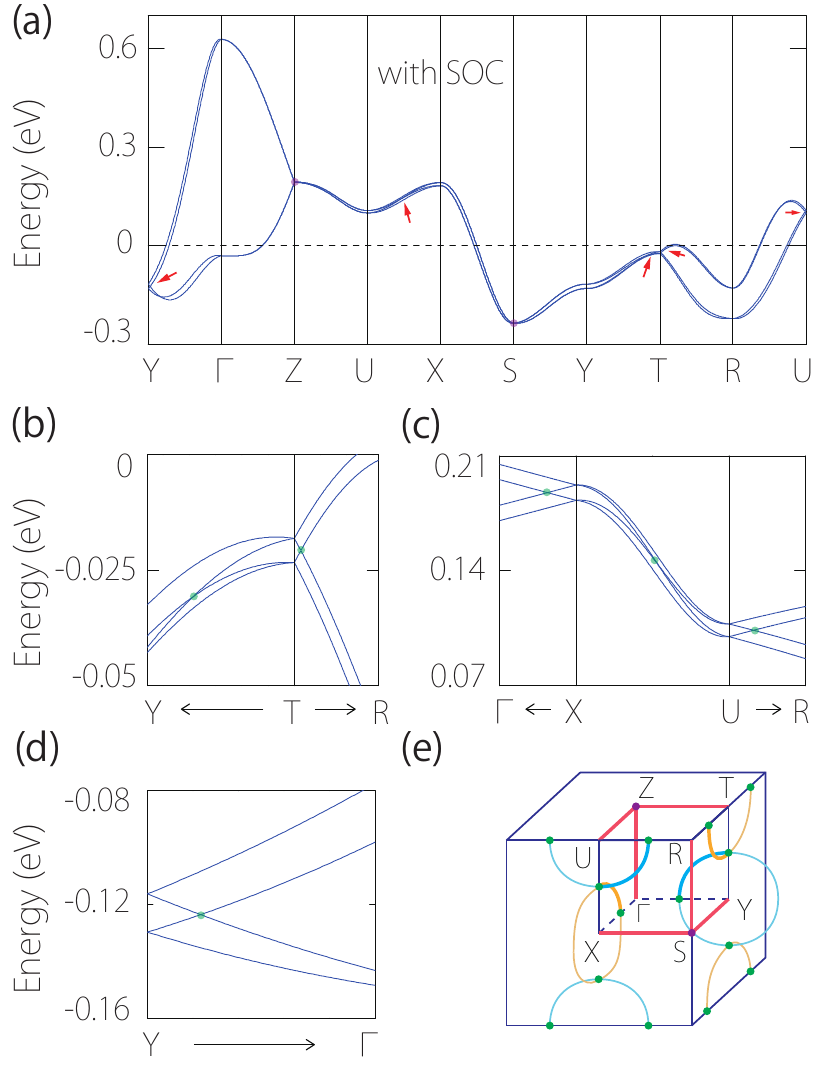}
\caption{(a) Low-energy band structure of Li$_3$(TeO$_3$)$_2$. SOC is included. (b)-(d) Enlarged views of the band structure around the hourglass crossing points. (e) Schematic figure showing the Weyl chains in the 3D BZ. The blue and yellow lines show the Weyl loops located in mutually orthogonal planes. The high-symmetry lines supporting a twofold degeneracy are highlighted in red. }
\label{TeSOCband}
\end{figure}

The band structure shown in Fig.~\ref{TeSOCband}(a) has the following features: (i) The bands along $\Gamma$-$Z$, $R$-$S$, $U$-$Z$, $Z$-$T$, $X$-$S$, and $S$-$Y$ are doubly degenerate; (ii) hourglass dispersions appear on several paths, such as $Y$-$T$, $T$-$R$, $\Gamma$-$X$, $U$-$R$, and $Y$-$\Gamma$.

\begin{figure}[t!]
\includegraphics[width=8.5 cm]{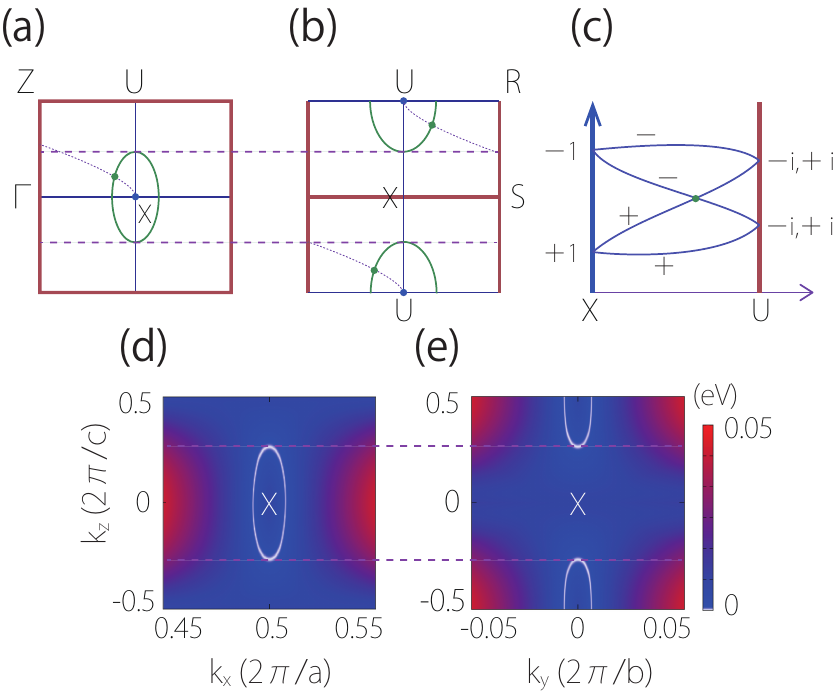}
\caption{Schematic figures showing that the neck point in the hourglass dispersion traces out Weyl loops on (a) $k_y=0$ plane and (b) $k_x=\pi$ plane. (c) Schematic view of hourglass dispersion along $X$-$U$. The labels indicate the eigenvalues of   $\mathcal{\tilde{M}}_x$. (d)(e) Shape of the Weyl loops from the DFT calculations, (d) on the $k_y=0$ plane and (e) on the $k_x=\pi$ plane, respectively. The color map indicates the local gap between the two crossing bands. }
\label{chain}
\end{figure}

First, we show the double degeneracy on those high symmetry paths are guaranteed by symmetry. The paths $\Gamma$-$Z$ and $S$-$R$ are invariant subspaces for $\mathcal{\tilde{M}}_x$ and $\mathcal{\tilde{M}}_y$. In the presence of SOC, we have
\begin{equation}
\mathcal{\tilde{M}}_x \mathcal{\tilde{M}}_y = \overline{E} T_{\bar{1}10} \mathcal{\tilde{M}}_y \mathcal{\tilde{M}}_x,
\end{equation}
where $\overline{E}=-1$ comes from the anti-commutativity between two spin rotations, i.e., $\{\sigma_x,\sigma_y\}=0$. For paths $\Gamma$-$Z$ and $S$-$R$, $k_y - k_x = 0$, $\overline{E} T_{\bar{1}10}=-1$, so $\mathcal{\tilde{M}}_x$ and $\mathcal{\tilde{M}}_y$ anti-commute. Similar to the arguments after Eq.~(\ref{Eq1}), this anti-commutation indicates that all bands along $\Gamma$-$Z$ and $S$-$R$ are doubly degenerate.

Meanwhile, the double degeneracy along $U$-$Z$, $Z$-$T$, $X$-$S$, and $S$-$Y$ are enabled by the glide mirror and $\mathcal{T}$ symmetries. Here, we take the $U$-$Z$ path ($k_y=0$ and $k_z=\pi$) as an example. We have
\begin{equation}
(\mathcal{T\tilde{M}}_x )^{2} =  -\overline{E}T_{011}= e^{-i k_y - i k_z},
\end{equation}
where we have used $\mathcal{T}^2=-1$. Thus, $(\mathcal{\tilde{M}}_x \mathcal{T})^2=-1$ on $U$-$Z$, indicating a Kramers-like double degeneracy.  Similar analysis applies for the other three paths $Z$-$T$, $X$-$S$, and $S$-$Y$.

Next, we turn to the hourglass dispersion. Consider the $X$-$U$ line, which is invariant under $\mathcal{\tilde{M}}_y $. Hence, the Bloch states there can be chosen as eigenstates of $\mathcal{\tilde{M}}_y $ with eigenvalues
\begin{equation}
g_y = \pm e^{-i k_z/2}.
\end{equation}
The glide eigenvalues are $\pm i$ at $U$ and $\pm 1$ at $X$. Because $U$ and $X$ are both time reversal invariant momenta, a Kramers pair has eigenvalues $(+i,-i)$ at $U$, yet it has $(+1,+1)$ or $(-1,-1)$ at $X$. This means that the pairs must switch partners when going from $X$ to $U$, and the switching leads to the hourglass type dispersion, as shown in Fig.~\ref{chain}(c).

Since the whole $k_y=0$ plane is invariant under $\mathcal{\tilde{M}}_y$, the above argument holds for any in-plane path connecting $X$ and $U$.
In addition, as mentioned above, bands along $\Gamma$-$Z$ and $Z$-$U$ are all doubly degenerated, so any path connecting $X$ to an arbitrary point $P$ located on $\Gamma$-$Z$ or $Z$-$U$ should also has the hourglass dispersion. The neck point of the hourglass then traces out a Weyl loop surrounding $X$, as shown in Fig.~\ref{chain}(a).

Similar analysis applies for the $k_x=\pi$ planes, if $\mathcal{\tilde{M}}_y $ is replaced by  $\mathcal{\tilde{M}}_x$. This gives another Weyl loop surrounding $U$ [Fig.~\ref{chain}(b)]. The two loops are orthogonal to each other and touch at a point on $U$-$X$, thus forming a Weyl chain as shown in Fig.~\ref{TeSOCband}(e). Likewise, one can show that there exists another chain running along $T$-$Y$, as in Fig.~\ref{TeSOCband}(e).

\end{appendix}

\bibliography{LiFeO_refv2}

\end{document}